\documentclass{Interspeech2024}

\interspeechcameraready 
\usepackage{times}
\usepackage{latexsym}
\usepackage{soul}
\usepackage[T1]{fontenc}
\usepackage{verbatim}
\usepackage{booktabs}
\usepackage{hyperref}
\usepackage{multirow}
\usepackage{graphicx}
\usepackage{amsmath}
\usepackage{subcaption}
\usepackage[most]{tcolorbox}

\usepackage[utf8]{inputenc}

\usepackage{microtype}

\usepackage{inconsolata}
\usepackage{graphicx}
\usepackage{subcaption}

\title{ComFeAT: Combination of Neural and Spectral Features for Improved Depression Detection}

\name[affiliation={1}]{Orchid}{Chetia Phukan*}
\name[affiliation={2}]{Sarthak}{Jain*}
\name[affiliation={3}]{Shubham}{Singh*}
\name[affiliation={4}]{Muskaan}{Singh}
\name[affiliation={1}]{Arun Balaji}{Buduru}
\name[affiliation={1,5}]{Rajesh}{Sharma}

\address{
  $^1$IIIT-Delhi, India,
  $^2$GGSIPU, New Delhi, India,
  $^3$BIT Mesra, India,
  $^4$ISRC, Ulster University, UK\\
  $^5$University of Tartu, Estonia, *equal contribution}
\email{orchidp@iiitd.ac.in, sarthakjainssjj@gmail.com, shubhamsingh051222002@gmail.com}


\begin{document}

\maketitle

\begin{abstract}
In this work, we focus on the detection of depression through speech analysis. Previous research has widely explored features extracted from pre-trained models (PTMs) primarily trained for paralinguistic tasks. Although these features have led to sufficient advances in speech-based depression detection, their performance declines in real-world settings. To address this, in this paper, we introduce \textbf{ComFeAT}, an application that employs a CNN model trained on a combination of features extracted from PTMs, a.k.a. neural features and spectral features to enhance depression detection. Spectral features are robust to domain variations, but, they are not as good as neural features in performance, suprisingly, combining them shows complementary behavior and improves over both neural and spectral features individually. The proposed method also improves over previous state-of-the-art (SOTA) works on E-DAIC benchmark.  
\end{abstract}

\section{Introduction}
Depression is a prevalent mental health disorder that affects more than 280 million people world-wide. It is often characterized by persistent sadness, loss of interest, and other symptoms that impact daily functioning \footnote{https://www.who.int/news-room/fact-sheets/detail/depression}. Depression can manifest uniquely in each individual, sometimes alongside other mental health conditions or can be the cause of other mental heatlh condition such as anxiety. Early intervention can significantly help in improving outcomes and restoring the quality of life for those affected by depression. Recent studies in the field of speech processing and ML have shown promising results in automatically identifying various speech characteristics that are associated with depression such as low pitch, reduced pitch variability, slower speaking rate, increased pause frequency, and changes in voice quality. 
\par
Previous research have explored various features for effective depression detection ranging from spectral features such as MFCC \cite{das2024deep} also features from pre-trained models (PTMs) \cite{flores2021depression, brueckner2024audio} i.e neural features. 
Amongst the features, features from PTMs specifically trained for paralinguistic tasks such as speaker recognition, speech emotion recognition, and so on have shown better performance, and widely preferred \cite{campbell23_interspeech, brueckner2024audio}. However, they still lack in performance in real-world settings due to various variances present in real-world. In this demonstration, we present, \textbf{ComFeAT}, an application that makes use of CNN-based model trained with combination of neural and spectral features for improved depression detection in real-world. We experiment with different state-of-the-art (SOTA) spectral and neural features with CNN as the modeling paradigm and the topmost performance is shown by models trained on combination of neural and spectral features. This combination methodology also improves over previous SOTA works on E-DAIC benchmark.

\begin{figure}[hbt!]    
\centering
      \includegraphics[width=0.36\textwidth, height=0.45\textwidth]{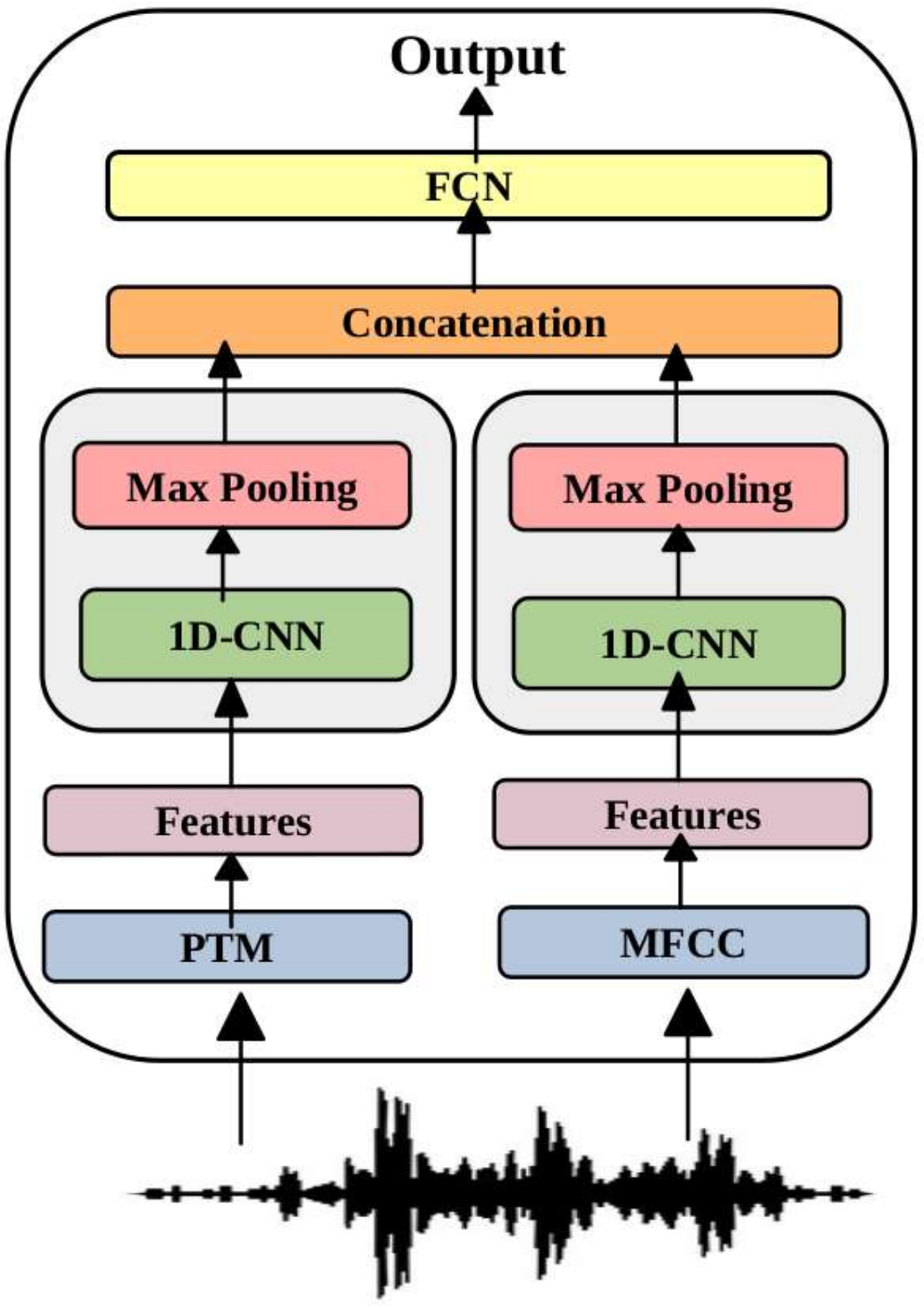}
      \caption{Proposed Model Architecture}
        \label{fig:archi}     
\end{figure} 

\vspace{-0.5cm}
\section{ComFeAT}

In this section, we discuss the various parts of \textbf{ComFeAT} application. First, we discuss the features considered in our experiments, followed by the classifier, database used, and evaluation results. Lastly, we discuss \textbf{ComFeAT} application buildup, workflow and the user interface.  


\noindent\textbf{Features}: We use MFCC and LFCC as spectral features. For neural features, we choose, TRILLsson and x-vector which has been commonly used in previous works \cite{campbell23_interspeech, brueckner2024audio}. TRILLsson \cite{shor22_interspeech} is distilled version of paralinguistic conformer which has been trained primarily for paralinguistic tasks while x-vector for speaker recognition. TRILLsson has shown SOTA performance in various paralinguistic tasks, and x-vector in speaker recognition. We extract features of 1024, 512 dimension size from TRILLsson and x-vector respectively, through averaging across time.  \par
\noindent\textbf{Modeling}: We use CNN for our experiments with different features. For experimentation with individual features we use 1D-CNN after the features followed by max-pooling and a fully-connected network (FCN) with two dense layers of 256, 90 neurons. Lastly, the output layer with mean-squared-error (MSE) as loss function. We use Adam as optimizer and we employ dropout to reduce overfitting. For experiments with combination of features, we employ the same modeling pattern with the 1D-CNN, max-pooling and concatenation for feature fusion followed by FCN and it is shown in Figure \ref{fig:archi}. We use 32 filters with size 3 in the 1D-CNN layer.  \par

\noindent\textbf{Database and Data Preprocessing}: We use the EDAIC dataset \cite{gratch2014distress} comprising audio recordings of individuals of varying emotional state, including depression. It contains more than 500 participants annotated with depression severity levels and demographic information. We resample the audio to 16KHz before feature extraction.

\begin{figure}[hbt!]    
\centering
      \includegraphics[width=0.52\textwidth, height=0.44\textwidth]{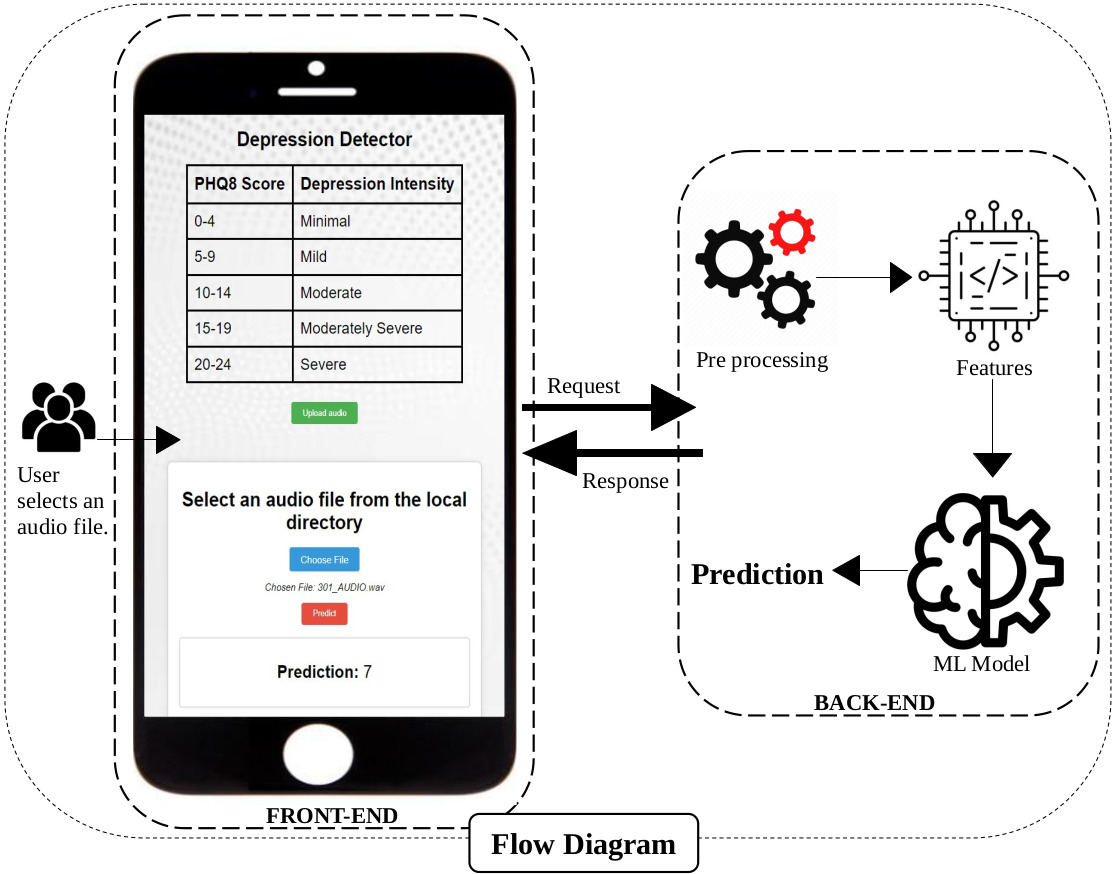}
      \caption{Flow Diagram}
        \label{fig:flow}     
\end{figure}

\begin{table}[hbt!]
\centering
\caption{Evaluation Scores; Lower the MAE and RMSE better the model performance}
\label{tab:embed}
\begin{tabular}{lll}   
\toprule
\textbf{Input Features} & \textbf{MAE} & \textbf{RMSE}\\
\midrule
LFCC & 6.02 & 7.07\\
MFCC & 5.35 & 6.71\\ 
x-vector & 5.03 & 6.32\\
TRILLsson & 4.48 & 5.83\\ 
x-vector + MFCC & 4.39 & 5.66\\
TRILLsson + MFCC & \textbf{4.05} & 5.57\\
x-vector + LFCC & 4.55 & 5.57\\
TRILLsson + LFCC & 4.23 & \textbf{5.28}\\
SOTA \cite{ringeval2019avec} & --- & 8.19\\
\bottomrule
\end{tabular}
\end{table}

\noindent\textbf{Evaluation Results}: We present our results in Table \ref{tab:embed} for different input features. Among the individual features, TRILLsson features attained the topmost performance. Models on combination of TRILLsson + MFCC and TRILLsson + LFCC attained the topmost performance amongst all the models in terms of MAE and RMSE, thus, showing the superiorty in combining neural and spectral features. We also achieve SOTA performance in terms of RMSE in comparison to previous SOTA work.  \par

\section{ComFeAT: Buildup and User Interface}

In this section, we present our \textbf{ComFeAT} buildup, workflow, and user interface (UI) in Figure \ref{fig:flow}. \par
\noindent\textbf{Buildup}: The application front-end is built using ReactJS, and the back-end is built using Flask. Flask is used for exposing the API of the application, where information to and fro from the proposed model flows. \par
\noindent\textbf{Workflow and User Interface}:
The workflow of \textbf{ComFeAT} is shown in Figure \ref{fig:flow}. A raw audio file in .wav, .mp3 etc is uploaded by a user via the upload audio button. This request is sent to the backend where the audio file is further pre-processed before passing it to the model and the preprocessing steps include resampling to 16kHz, converting to uni channel (stereo to mono if not), and extracting features. Now, the model makes predictions on the input features received. The predicted output is returned with a POST request to the frontend (UI). From the predicted value, the user can differentiate the intensity of depression. 

\section{Conclusion}
In this demonstration, we will present, \textbf{ComFeAT}, for real-world depression detection. We leverage the combination of neural and spectral features for improved depression detection than previous works. \textbf{ComFeAT} will find applications in various domains such as assisting mental health professionals, educational institutions and student support centers, community mental health support centers, and many more. 
\bibliographystyle{IEEEtran}
\bibliography{main.bib}

\end{document}